\documentclass{article}
\usepackage{graphicx}
\usepackage{amsmath}

\begin{document}

\pagestyle{empty} 
\begin{flushright}
{CERN-TH/2001-128}\\
hep-ph/0105168\\
\end{flushright}
\vspace*{15mm}

\begin{center}
\textbf{ A NEW SUM RULE TO DETERMINE $|V_{ub}|/|V_{cb}|$ } \\[0pt]
\vspace*{1cm} \textbf{U. Aglietti}$^{\ast )}$ \\[0pt]
\vspace{0.3cm} Theoretical Physics Division, CERN\\[1pt]
CH - 1211 Geneva 23 \\[3pt]
$~~~$ \\[0pt]
\vspace*{2cm} \textbf{Abstract} \\[0pt]
\end{center}

We present a new sum rule which does not contain any
Landau-pole effect and allows a determination of
$|V_{ub}|/|V_{cb}|$ with a
theoretical error of $O\left(5\%\right).$
\vspace*{4cm}

\vfill
\noindent Invited talk at the LEPTRE Conference, Rome 18-20 April 2001, 
to be published in the proceedings

\vspace*{0.3cm}

\noindent 
\rule[0.1in]{12cm}{0.002in}

\noindent 
$^{\ast )}$ On leave of absence from Dipartimento di Fisica,
Universita' di Roma I, Piazzale Aldo Moro 2, 00185 Roma, Italy. E-mail
address: ugo.aglietti@cern.ch. 

\begin{flushleft} CERN-TH/2001-128 \\[0pt]
May 2001
\end{flushleft}
\vfill\eject

\setcounter{page}{1} \pagestyle{plain}

\section{ Introduction}

Semileptonic and radiative $B$-decays, 
\begin{equation}
B\rightarrow X_{u}+l+\nu \qquad \mathrm{and\qquad }B\rightarrow X_{s}+\gamma
,  \label{unodue}
\end{equation}
are affected by a variety of non-perturbative effects:

\begin{enumerate}
\item  \textit{Beauty mass entering all decay rates.} As well known,
inclusive as well as semi-inclusive rates are proportional to the fifth
power of the $b$-quark --- not $B$-meson --- mass, because the elementary
process in (\ref{unodue}) is 
\begin{equation}
b\rightarrow \widehat{X}_{q}+\left( \mathrm{non\,\,QCD\,\,partons}\right) .
\label{partonico}
\end{equation}
There is a non-perturbative splitting of the initial energy $\mathcal{E}%
=m_{B}$ coming from the $B$-mass into a fraction $m_{b}$ partecipating to
the hard process and a remaining part $m_{B}-m_{b}$ acting as spectator.
This problem is usually solved by taking ratios of distributions, in which
the $m_{b}^{5}$ dependence cancels out. Furthermore, the HQET allows to
consistently include the non-perturbative corrections to branching fractions
which are suppressed by inverse powers of $m_{b}.$

\item  \textit{Fermi-motion.} Dynamics become more complicated going from
inclusive rates to distributions in the threshold region 
\begin{equation}
m^{2}\ll Q^{2},  \label{threshold}
\end{equation}
where $m$ is the jet mass and $Q=2E\gg \Lambda $ is the \textit{hard scale } 
\cite{ultimo}, with $E$ the jet energy in the $B$ rest-frame and $\Lambda $
the QCD scale. There are various regimes satisfying condition (\ref
{threshold}): $i)$\thinspace \thinspace the region 
\begin{equation}
\Lambda Q\ll m^{2}\ll Q^{2},  \label{ancorapert}
\end{equation}
is characterized by the occurrence of double logarithms $\alpha _{S}\left(
Q\right) ^{n}\log ^{2n}\left( 1-z\right) ,$ where $z\equiv 1-m^{2}/Q^{2}$ is
the fundamental variable in semi-inclusive heavy flavour decays. To have a
consistent result, these large logarithms need to be resummed to all
orders
in perturbation theory. Region (\ref{ancorapert}) is then described by
resummed perturbation theory. Let us observe that, in the real world, the
window (\ref{ancorapert}) is quite small \cite{noi3}; $ii)\,\,$if one pushes
the jet mass to even smaller values, in the slice 
\begin{equation}
m^{2}\sim \Lambda \,Q,  \label{unmezzo}
\end{equation}
resummation of infrared logarithms is no longer sufficient\footnote{%
Actually, the resummed distribution becomes singular in region (\ref{unmezzo}%
) because of Landau pole effects.} since new non-perturbative phenomena come
into play, the well-known Fermi motion effects \cite{nostri}. The exchange
of soft momenta between the beauty quark and the light valence quark in the $%
B$-meson cannot be neglected. In sec.\thinspace 3 we show that it is
possible to construct ratios of distributions in which the Fermi-motion
effects completely cancel.

\item  \textit{Final state hadronization.} If the jet mass is pushed to even
smaller values, $m^{2}\sim \Lambda ^{2},$ the exclusive or resonance region
is encountered. Additional non-perturbative effects come into play, related
to the hadronization of the partons in the final state.
\end{enumerate}

\section{Fermi motion effects, universality and factorization}

In the rest of this talk, we concentrate on the non-perturbative phenomena
occurring in region (\ref{unmezzo}). In agreement with classical intuition 
\cite{nostri}, Fermi motion effects have universality properties which allow
one to write general factorization formulae. Actually, in QCD, this follows
from the universal properties of soft and collinear radiation. The photon
spectrum in the rare decay in (\ref{unodue}) can be written as \cite{ultimo}%
: 
\begin{equation}
\frac{d\Gamma _{rd}}{dx}=C_{rd}\left( \alpha _{S}\right) \,f\left( x;\alpha
_{S}\right) +D_{rd}\left( x;\alpha _{S}\right) ,  \label{fac1}
\end{equation}
where $x\equiv 2E_{\gamma }/m_{B}.$ The coefficient function $C_{rd}\left(
\alpha _{S}\right) $ and the remainder function $D_{rd}\left( x;\alpha
_{S}\right) $ are process-dependent and short-distance dominated; they can
therefore be safely computed in fixed-order perturbation theory: 
\begin{equation}
C_{rd}\left( \alpha _{S}\right) \,\,\,=c_{rd}^{0}+\frac{\alpha _{S}C_{F}}{%
\pi }c_{rd}+\cdots ;\qquad D_{rd}\left( x;\alpha _{S}\right) =\frac{\alpha
_{S}C_{F}}{\pi }d_{rd}\left( x\right) +\cdots .
\end{equation}
$D_{rd}$ starts at order $\alpha _{S}$ (the explicit expressions of these
functions can be found in \cite{ultimo}). The function $f\left( x;\alpha
_{S}\right) $ contains all the non-perturbative effects related to Fermi
motion and is universal, i.e. process-independent\footnote{%
The function $f$ does not exactly coincide with the shape function $f^{ET}$
defined in the low-energy effective theory, because it also contains some
short-distance effects.}. Condition (\ref{unmezzo}) corresponds to $x\sim
1-\Lambda /m_{B}\sim 0.9\div 0.95\footnote{%
In the region $x\ll 1-\Lambda /m_{B}$ the function $f$ is perturbative, as
this corresponds to condition (\ref{ancorapert}).}.$ Integrating on both
sides of eq.\thinspace (\ref{fac1}) over $x$, the total rate is correctly
reproduced. A measure of the photon spectrum allows a direct determination
of $f.$ A similar factorization formula can also be written for the
triple-differential distribution in the semi-leptonic decay (\ref{unodue}) 
\cite{ultimo}: 
\begin{equation}
\frac{d^{3}\Gamma _{sl}}{dx_{e}dwdz}=C_{sl}\left( w,x_{e};\alpha _{S}\right)
\,f\left( z;\alpha _{S}\right) +D_{sl}\left( w,x_{e},z;\alpha _{S}\right) ,
\label{tripla}
\end{equation}
where $x_{e}\equiv 2E_{e}/m_{B}$ and $w\equiv Q/m_{B}.$ The coefficient
function now depends also on $x_{e}$ and $w:$ the main point however is that 
$C_{sl}$ does not depend on the ``Fermi motion variable'' $z.$
Equation\thinspace (\ref{tripla}), from which any other distribution is
obtained by integration, then describes the effects of Fermi motion in any
semileptonic decay spectrum. An interesting quantity is the hadron energy
spectrum $d\Gamma _{sl}/dE$, whose behaviour for $E=m_{B}/2+O\left( \Lambda
\right) $ is non-perturbative and is controlled by the cumulative
distribution $F$ (the integral of  $f).$ A measure of $d\Gamma _{sl}/dE$ in
this region would allow an independent determination of $f$ \cite{ultimo}.
However, the most promising distribution to determine $f$ is that in the $z$%
-variable. Integrating over the energies $w$ and $x_{e},$ one obtains: 
\begin{equation}
\frac{d\Gamma _{sl}}{dz}=C_{sl}\left( \alpha _{S}\right) \,f\left( z;\alpha
_{S}\right) +D_{sl}\left( z;\alpha _{S}\right) 
\end{equation}
where 
\begin{equation}
\frac{C_{sl}\left( \alpha _{S}\right) }{\Gamma _{sl}^{0}}=1+\frac{\alpha
_{S}C_{F}}{\pi }\left( \frac{115}{144}-\frac{\pi ^{2}}{2}\right) 
\end{equation}
and 
\begin{eqnarray}\nonumber
\frac{d_{sl}\left( z\right) }{\Gamma _{sl}^{0}}&=&\frac{8}{35}\,\frac{%
80+145t+44t^{2}+11t^{3}}{\left( 1+t\right) ^{5}\left( 1-t\right) }\log \frac{%
1+t}{1-t}\,+\frac{4}{1-t^{2}}\log \frac{1-t^{2}}{4}+
\\  
&&\quad\quad\quad\quad\quad\quad+\frac{%
21+t-49t^{2}-21t^{3}}{3\left( 1+t\right) ^{5}}, 
\end{eqnarray}
with $t\equiv \sqrt{1-4\left( 1-z\right) }=\sqrt{1-m^2/E^2}$ and $\Gamma
_{sl}^{0}=G_{F}^{2}\,m_{b}^{5}|V_{ub}|^{2}/(192\pi ^{3}).$ A measure of this
distribution then would provide still another direct and independent
determination of $f$\footnote{%
A measure of the distribution $d\Gamma _{sl}/dz$ is not available at present
and a problem for the data analysis may be the large $b\rightarrow cl\nu $
background. It is however possible to impose the usual kinematical cut $%
m<m_{D}$ in $d\Gamma _{sl}/dz$ without changing the infrared structure \cite
{noi3}.}. \ The idea underlying the unifying representations (\ref{fac1})
and
(\ref{tripla}) is that kinematics is different in the two processes, but
dynamics is the same.

\section{A new sum rule to determine $|V_{ub}|/|V_{cb}|$}

From the above discussion, it is clear that ratios of distributions can be
constructed which do not depend on $f$ and are therefore short-distance
quantities. Consider indeed: 
\begin{equation}
R\equiv \left. \frac{d\Gamma _{rd}/dx-D_{rd}\left( x;\alpha _{S}\right) }{%
d\Gamma _{sl}/dz-D_{sl}\left( z;\alpha _{S}\right) }\right|
_{z=x\geq 0.75}=\frac{%
C_{rd}\left( \alpha _{S}\right) }{C_{sl}\left( \alpha _{S}\right) }.
\label{forsenuova}
\end{equation}
Non-perturbative effects completely cancel and there is consequently no
Landau-pole singularity in $R$ for any value of $z.$ The sum rule
represented by eq.\thinspace (\ref{forsenuova}) is, as far as we know, new.
Different sum rules for the determination of $|V_{ub}|/|V_{cb}|$ ---
involving the jet-mass distribution $d\Gamma_{sl}/dm$ or the
electron spectrum $d\Gamma_{sl}/dx_e$ instead of $d\Gamma
_{sl}/dz$ --- have also been proposed in \cite{rozzo} and \cite{3tipi}.
Other sum rules can be obtained by considering the hadron energy
distribution. Considering for illustrative purposes the operator $O_{7}$
only, for which 
\begin{equation}
\frac{C_{rd}}{\Gamma _{rd}^{0}}=1-\frac{\alpha _{S}C_{F}}{\pi }\left( \frac{%
\pi ^{2}}{3}+\frac{13}{4}\right) ,~ \frac{d_{rd}\left( x\right) }{%
\Gamma _{rd}^{0}}=\frac{1}{4}\left[ 7+x-2x^{2}-2\left( 1+x\right) \log
\left( 1-x\right) \right] ,
\end{equation}
with $\Gamma _{rd}^{0}=\alpha _{em}G_{F}^{2}\,m_{b}^{5}|V_{tb}V_{ts}^{\ast
}|^{2} C_7^2(m_b)  /(32\pi ^{4}),$ the ratio $R$ reads\footnote{%
A more complete analysis is in progress \cite{noi3}.}: 
\begin{equation}
R\simeq \frac{6\alpha _{em}}{\pi }C_{7}^2\left( m_{b}\right) \left[
1+%
\frac{\alpha _{S}C_{F}}{\pi }\left( \frac{\pi ^{2}}{6}-\frac{583}{144}%
\right) \right] \frac{|V_{cb}|^{2}}{|V_{ub}|^{2}}\simeq 1.05\cdot 10^{-3}%
\frac{|V_{cb}|^{2}}{|V_{ub}|^{2}},
\end{equation}
where we have taken $V_{tb}\simeq 1,V_{ts}\simeq V_{cb},$ $C_{7}\left(
m_{b}\right) \simeq -0.31$ and $\alpha _{S}=\alpha _{S}\left( m_{b}\right)
\simeq 0.21$ \cite{citare}. A comparison of \ a more complete expression
for $%
R$ with the data should allow an accurate determination of $|V_{ub}|/|V_{cb}|
$ \cite{noi3}. The ``irreducible'' theoretical error comes from the
higher-twists: $\delta |V_{ub}|^{2}/|V_{cb}|^{2}$ $\sim \Lambda
/m_{B}\sim 10\%,$ so that $\delta |V_{ub}|/|V_{cb}|$ $\sim O\left(
5\%\right) .$

\section{Conclusions}

Fermi motion effects in semileptonic and rare $B$ decays can be factorized
by means of a universal non-perturbative function $f\left( z\right) .$ The
latter can be determined by measuring the photon spectrum in the rare decay (%
\ref{unodue}) in the end-point region. In the semileptonic case, any
distribution can in principle be used for the experimental determination of $%
f.$ Two distributions are particularly interesting: the hadron energy
spectrum in the region $E=m_{B}/2+O\left( \Lambda \right) $ and the $z$%
-distribution for $z\sim 1-\Lambda /m_{B},$ as they are locally proportional%
\footnote{%
By this we mean that no convolution of $f$ with some weight function is
involved.} to $f$ in the non-perturbative region. \ A sum rule was presented
which allows a model-independent extraction of \ $|V_{ub}|/|V_{cb}|$ with a
theoretical error of $O\left( 5\%\right) .$

\end{document}